\documentclass[aps,prl,twocolumn,preprintnumbers,showpacs,amsmath,amssymb,nobalancelastpage]
{revtex4}
\usepackage{amsfonts,mathrsfs,graphicx}
\usepackage{epsfig}
\usepackage{bm}
\usepackage{color}
\def\be{\begin{equation}}
\def\ee{\end{equation}}
\def\bea{\begin{eqnarray}}
\def\eea{\end{eqnarray}}
\def\nnb{\nonumber}

\begin{document}

\title{Scanning the Earth with solar neutrinos and DUNE}

\author{
 A. N. Ioannisian$^{1,2}$, A.Yu. Smirnov$^{3,4}$, D. Wyler$^5$
 }
 \address{
 $^1$
Yerevan Physics Institute, Alikhanian Br.\ 2, 375036 Yerevan,
Armenia\\
$^2$Institute for Theoretical Physics and Modeling, 375036
Yerevan, Armenia\\
$^3$
Max-Planck Institute for Nuclear Physics, Saupfercheckweg 1, D-69117 Heidelberg, Germany \\
$^4$
ICTP, Strada Costiera 11, 34014 Trieste, Italy \\
$^5$
Institut f\"ur Theoretische Physik, Universit\"at Z\"urich,
Winterthurerstrasse 190, CH-8057 Z\"urich, Switzerland
}

\begin{abstract}
We explore oscillations of the solar $^8$B neutrinos in the Earth in detail.  
The relative excess of night $\nu_e$ events (the Night-Day asymmetry) 
is computed as function of the 
neutrino energy and the nadir angle $\eta$ of its trajectory. 
The finite energy resolution of  
the detector causes an important attenuation effect, while the layer-like structure of the Earth density
leads to an interesting parametric 
suppression of the oscillations. 
Different features of the $\eta-$ dependence 
encode information about the structure (such as density jumps) of the Earth density profile; thus
measuring the $\eta$ distribution  allows the scanning of the interior of the Earth.  
We estimate the sensitivity of the  DUNE experiment to such measurements.  
About 75 neutrino events are expected  per day in  40 kt.  
For high values of $\Delta m^2_{21}$ and $E_\nu > $11 MeV, the corresponding D-N asymmetry
is about 4\% and can be measured with $15\%$ accuracy
after 5 years of data taking. 
The difference of  the D-N asymmetry between high and low values of $\Delta m^2_{21}$ 
can be measured at the $4\sigma$ level. 
The relative excess of the $\nu_e$ signal  
varies with the nadir angle up to  50\%. DUNE may establish the existence of 
the dip in the $\eta-$ distribution at the $(2 - 3) \sigma$ level.
 
\end{abstract}

\preprint{CERN-TH-2016-256}


\maketitle

\section{I  \ Introduction}

The Earth matter effect on solar neutrinos predicted long time ago 
\cite{ms4, Carlson:1986ui, Cribier:1986ak, Bouchez:1986kb, Hiroi:1987pe, 
Baltz:1986hn, Dar:1986pj, ms87, Baltz:1988sv} 
has been succesfully established at more than $3\sigma$ level \cite{Renshaw:2013dzu}, \cite{Abe:2016nxk}. 
SuperKamiokande (SK) has observed the Day-Night (D-N) asymmetry of the solar neutrino 
signal defined as $A_{DN}^{s}  \equiv  2 (N_D - N_N)/ (N_N + N_D)$, 
where $N_N$ and $N_D$ are the rates of events 
detected during night and day. 
$A_{DN}$ has been determined  by the fit of the zenith angle dependence 
of $N_N$ and $N_D$ integrated over the energy interval 
(4.5 - 19.5) MeV. In turn, the rates $N_N$ and $N_D$ as functions 
of zenith angle (shapes) were computed according to the LMA MSW solution 
for certain values of oscillation parameters. The amplitude of the D-N variations 
was used as fit parameter. 
In this way, using the best value from the solar data fit (low value), 
$\Delta m^2_{21} = 4.8 \cdot 10^{-5} $eV$^2$, the asymmetry 
\be
A_{DN}^{s, fit} = - [3.3 \pm 1.0 (stat.)\pm 0.5 (syst.)] \%
\label{eq:asym}
\ee
has been found from a combination of  all series of measurements (SK I-IV). 

Assuming that there is no energy and zenith angle dependences of the 
oscillation effect, SK finds a larger asymmetry: 
$A_{DN}^s = - [4.2 \pm 1.2 (stat.)\pm 0.8 (syst.)]$\%  (SK I-IV) and 
$A_{DN}^s = - [4.9 \pm 1.8 (stat.)\pm 1.4 (syst.)]$\%  in SK IV alone 
with additional statistics \cite{Renshaw:2013dzu}, \cite{Abe:2016nxk}.

Previously,  SNO measured the neutrino spectra during day and night,
indicating  a non-zero day-night asymmetry  
\cite{Ahmad:2002ka}, \cite{Aharmim:2005gt}.  
Joint analysis  of all solar neutrino data gives about a
$4\sigma$ evidence of the Earth matter effect \cite{Maltoni:2015kca}.  
SuperKamiokande  also presented the first meaningful measurements 
of the zenith angle distribution of events.  

These results mark the beginning of experimental exploration 
of the Earth matter effects. 
Future  developments in detection techniques and the construction 
of large mass detectors with high energy resolution will  open up the possibility
of detailed study of the solar neutrino oscillations 
in the Earth. The implications of these studies include 
tomography (scanning) of 
interior of the Earth, measurements of the  neutrino parameters and searches for  
new physics beyond the $3\nu-$paradigm.

In fact, there are  some hints of new physics already now: 
the result (\ref{eq:asym}) has reinforced a  tension 
in determination of $\Delta m^2_{21}$. 
Indeed,  the asymmetry  (\ref{eq:asym}) is in  good agreement with 
the prediction from LMA MSW at  low value of $\Delta m^2_{21}$ 
which provides consistent description of all solar neutrino data. 
The asymmetry (\ref{eq:asym}) is about $2\sigma$ larger  than the asymmetry
$1.7\%$  expected for the global fit (high) value of $\Delta m^2_{21}$
dominated by the KamLAND \cite{Gando:2010aa} data.   
Apart from the large observed D-N asymmetry, also an absence of the 
``spectral upturn'' at low energies 
indicates a low value of $\Delta m^2_{21}$ \cite{Maltoni:2015kca}.  
In future JUNO \cite{An:2015jdp} will measure $\Delta m^2_{21}$ with very high precision.

The large DN-asymmetry can be (i) just a statistical fluctuation,  
(ii) a result of incorrect computation of the expected asymmetry or 
(iii) due to presence of large  (non-standard) matter effect. 
 
Recall that oscillations of solar neutrinos in the Earth,  
being  oscillations of the mass states,  
are  pure matter effect proportional 
to the matter potential $V$.  
In the case of standard neutrino interactions, a 
large $V$ is only possible if the chemical composition 
is abnormal since the density profile of the Earth is very well known 
\cite{Dziewonski:1981xy, shearer, petersen1993}. 
The standard matter effect is determined by the  
number density of electrons $n_e$ which is related to the total density $\rho$ as 
$n_e = Y_e \rho/m_N$. Here  $m_N$ is the nucleon mass and  $Y_e$  
is the electron fraction.  $Y_e = 1/2$ for an isotopically 
neutral medium (which is realized in the mantle) and $Y_e = 1$
for Hydrogen. Therefore the required increase 
of the matter effect is possible if one assumes an
abnormally high percentage of hydrogen in the Earth. 
According to the hydric  model of the Earth \cite{Bezrukov:2016gmy}, a large abundance of hydrogen
exists in the core of the Earth. However,  due to the attenuation effect to be discussed below,
SK (having bad reconstruction of the neutrino energy) 
is not sensitive to  the core. So,  
even this exotic model does not allow to resolve the tension.

Still another possibility is to assume the existence of non-standard 
neutrino interactions \cite{Friedland:2004pp},  \cite{Miranda:2004nb}
(see also the review \cite{Maltoni:2015kca}). We do not further elaborate on this.
 
In  anticipation of future experiments, 
in this paper we study in detail the Earth matter effect on the high energy 
part of the Boron neutrino spectrum. For previous studies see 
\cite{Ioannisian:2004jk, Akhmedov:2004rq, Ioannisian:2004vv} and references therein. 
(Oscillations of the solar $^7$Be neutrinos in the Earth have been explored  
in \cite{Ioannisian:2015qwa}.) 
Here e compute the nadir angle dependence 
($\eta \equiv \pi - \theta_z$,  $\theta_z$ being the zenith angle) of the relative  excess of the night events 
\begin{equation}
A_{ND} (\eta, E) \equiv \frac{N_N}{N_D} - 1 
\label{eq:excess}
\end{equation}
for different energies~\footnote{This definition differs 
from the standard D-N asymmetry:
$A_{DN}^{s} = - A_{ND} (1 + 0.5 A_{DN})^{-1}$, 
so  the one in (\ref{eq:excess}) is positive and about 2\% smaller.}.    
The distributions  $A_{DN} (\eta, E)$ is then integrated over energy, weighted
with the energy resolution (reconstruction) function of a detector.   
We study the dependence of these integrated distributions on the 
width of the energy resolution function. 
A complete interpretation of the nadir  angle distributions   
and their dependence on features of the Earth density profile is given. 
Thus, studies of the nadir angle distribution allow 
to scan the density profile of the Earth which is not possible for fixed 
$\eta$~ \cite{Akhmedov:2005yt}.

The paper is organized as follows: 
In Sec. II we summarize the relevant information on oscillations in the Earth. 
We further develop theory of neutrino oscillations in a multi-layer medium. 
The results are presented in the full three neutrino framework, see \cite{Blennow:2003xw}  
for the constant density case. In Sec. III 
we compute the relative excess $A_{ND}$ as function of the 
nadir angle and explore effect of the
integration over energy with an energy resolution function 
of different widths and give an interpretation of the obtained  dependences. 
As an example, in Sec. IV we estimate  the ability
of the DUNE experiment to measure the 
Earth matter effects. Conclusions are given in Sec. V.

\section{II \ Oscillations in the Earth}

Solar neutrinos arrive at the Earth 
as incoherent  fluxes of the mass eigenstates 
$\nu_i$. The fractions of these fluxes  are
determined by the mixing matrix elements in the production region: 
$P_{\nu_i} = |U_{ei}^m|^2$. 
In  the standard parametrization of the PMNS mixing matrix they equal  
\begin{equation}
P_{\nu_1}= c_{13}^2 \cos^2  \bar{\theta}^\odot_{12}, \ \ \ 
P_{\nu_2}= c_{13}^2 \sin^2  \bar{\theta}^\odot_{12}, \ \ \ 
P_{\nu_3} \approx s_{13}^2 , 
\end{equation} 
where the angle $\bar{\theta}^{\odot}_{12}$ is given by 
\begin{equation}
\cos 2\bar{\theta}^\odot_{12} \ \approx  \ 
{\cos 2 \theta_{12} - c_{13}^2 \bar{\epsilon}_\odot  
\over 
\sqrt{(\cos 2 \theta_{12} - c_{13}^2 \bar{\epsilon}_\odot)^2 + 
\sin^2 2 \theta_{12}}}. 
\label{cosi}
\end{equation}
Here the symbol $\odot$ refers to the solar production environment, 
\be
\label{epsi}
\epsilon_\odot   \equiv  \frac{2 V_\odot E}{\Delta m_{21}^2 },      
\ee
$c_{13} \equiv \cos \theta_{13}$, $s_{13} \equiv \sin \theta_{13}$.  
At low (solar neutrino) energies 
the matter effect on 1-3 mixing can be neglected,    
so that $ \bar\theta_{13}^\odot  \approx \theta_{13}=8.4^\circ$ \cite{An:2016ses}.  
Bars at $\theta$'s and $\epsilon$  mean averaging over the density 
in the $^8$B neutrino  production region.

In the Earth each mass state splits into 
eigenstates in matter 
and oscillates. 
Then the  probability to find $\nu_e$ in the detector equals 

$$
P = \sum_i P_{\nu_i} P_{ie} = \sum_i |U_{ei}^\odot |^2 P_{ie}, 
$$ 
where $P_{ie}$ is the probability of 
$\nu_i \rightarrow \nu_e$ transition in the Earth. 
The probability $P$ can be rewritten as 
\be
\label{probnue}
P =  c_{13}^2 (\cos 2 \bar{\theta}^\odot_{12} P_{1e} 
+ c_{13}^2 \sin^2 \bar{\theta}^\odot_{12})  + s_{13}^4,
\ee
where 
we used the unitarity 
relation: $P_{1e} + P_{2e} + s_{13}^2 = 1$ 
or $P_{2e} = c_{13}^2 - P_{1e}$.  Oscillations of $\nu_3$ are neglected.

In the Earth oscillations proceed in the low density regime when $\epsilon_m \ll 1$:  
at $E \sim 10$ MeV and surface density we have $\epsilon_m \sim 0.03$. Here,  
$\epsilon_m(x)$ is defined as in Eq. (\ref{epsi}), but with the potential $V_e$ taken 
in the Earth at position $x$;
a super(sub)script $m$ indicates that the respective quantity has to be taken 
inside the Earth matter.
Consequently, the oscillation length in matter  
\be
l_m = \frac{2\pi}{\Delta^m_{21} (x)} = l_\nu [1 + 
\cos 2\theta_{12} c_{13}^2 \epsilon_m(x) + O(\epsilon_m^2)], 
\label{osclength}
\ee
is rather close to the vacuum oscillation length 
\begin{equation}
l_m \approx l_\nu \approx 330 \ {\rm  km}
\left( \frac{7.5 \times 10^{-5}{\rm eV}^2}{\Delta m_{21}^2} \right) \! \! 
\left( \frac{E}{10 \ {\rm MeV}}  \right) \ .
\nnb
\end{equation}
In Eq. (\ref{osclength}) 
\begin{equation}
\Delta_{21}^m(x) \equiv \frac{\Delta m^2_{21}}{2E} \sqrt{[\cos 2
\theta_{12} - c_{13}^2\epsilon_m(x)]^2 + \sin^2 2 \theta_{12}} \ 
\label{split}
\end{equation}
is the splitting of the eigenvalues in matter at position x.

During day, $P_{1e} =  P^0_{1e}= c_{13}^2 \cos^2 \theta_{12}$     
and Eq. (\ref{probnue}) gives  
\begin{equation}
\label{PD}
P_D =  {c_{13}^4 \over 2 }(1  +  
\cos 2 \bar{\theta}^\odot_{12} \cos 2 \theta_{12} )  + s_{13}^4.  \  
\end{equation}
Then the total probability $P$ can be represented as
\begin{equation} 
\label{main1}
P  \equiv   P_D + \Delta P, 
\end{equation}
where
\begin{equation} 
\label{main3}
\Delta P \  =  
c_{13}^2 \cos 2 \bar{\theta}^\odot_{12}~ (P_{1e} - P^0_{1e})   
\end{equation}
describes the Earth matter effect. \\

The probability of the $\nu_1 \rightarrow \nu_e$ transition in the Earth,    
$P_{1e}$,  is determined by  dynamics of the $2\nu$-  
sub-system  in the propagation basis after decoupling of the third state
(see for details, e.g. \cite{Blennow:2003xw}, \cite{Maltoni:2015kca}).  
The propagation basis $\nu'$ is related to the original flavor basis $\nu_f$,  in particular,  
by the 1-3 rotation  $U_{13}(- \theta_{13})$.  The Hamiltonian of the $2\nu$ 
sub-system is characterized by the mixing angle  
$\theta_{12}$,  the mass squared difference $\Delta m^2_{21}$ 
and the potential $c_{13}^2 V_e$.   

The following derivation of  $P_{1e}$ reflects immediately the features 
of the density profile of the Earth which  can be considered as a multi-layer 
medium with slowly varying density inside the layers and sharp density changes  
(jumps) at the borders between the layers.  
Within the layers, due to  slow density change 
the neutrinos evolve adiabatically, that is,  
the transitions between the eigenstates are absent,  and they evolve independently. 
Indeed, departure from the adiabaticity is quantified by
the parameter $ \gamma$,
\be
\label{gomma}
\gamma \equiv \frac{1}{\Delta^m_{21}} \frac{d \theta^m_{12}}{dx} \approx 
\frac{\epsilon_m}{2\pi} \frac{l_m}{h_E},  
\ee
where $h_E \equiv V/(dV/dx)$ is the scale of the density change within  the layers.  
The second equality in (\ref{gomma}) follows from Eqs. (\ref{cosi})  and (\ref{epsi}) 
with  $\bar{\epsilon}_\odot$ substituted by $\epsilon_m$. $\theta^m_{12}$ is the mixing angle in matter.
To get an estimate of $\gamma$ we take $h_E = R_E$, where  $R_E$ is the radius of the Earth,  
and a typical oscillation length $l_m \approx l_\nu \approx 420$ km (for $E = 12.5$ MeV). 
This gives $\gamma =  1.6 \cdot 10^{-4}$, so that corrections to the adiabatic result 
are below $0.02\%$.  

At the borders of the layers, the adiabaticity in strongly (maximally) broken, 
which corresponds to a sudden change of     
the basis of eigenstates. Therefore after passing 
the border a different, coherent mixture of eigenstates emerges.

Within the layer the mixing angle $\theta^m_{12}$ changes slowly according to the density 
change. We will denote the values of the angle in the in the $k$th layer at its beginning 
and its end (along the neutrino trajectory) by $\theta_{12, k}^{m, i}$ and $\theta_{12, k}^{m, f}$.  

The transition matrix  between the initial mass states and final flavor states of the propagation basis,     
$\nu_i \to \nu_\alpha'$ in such a multi-layer medium  can be written  as 
\be  
S = U_n^m \Pi_{k = n, ... 1}   D_k  U_{k, k - 1}.   
\label{eq:ampl}
\ee
Here $n$ is the number of layers,  $U_n^m = U(\theta_{12, n}^m,f)$
is the flavor mixing matrix in the last layer just before a detector 
(it projects an evolved neutrino state  onto the flavor states), 
$\theta_{12, n}^m$ is the flavor mixing angle in the $n$-layer. 
The matrix of basis change between the $(k-1)$th and  $k$th layers 
\be
U_{k, k - 1} = U_{k, k - 1}(- \Delta \theta_{k-1}), 
\ee
is the matrix of rotation on the angle $\Delta \theta_{k-1}$, where  
\be
\Delta \theta_{k-1} \equiv \theta^{m,i}_{12, k}  - \theta^{m,f}_{12, k - 1} 
\label{eq:changeangle}
\ee
is the  difference of mixing angles in matter after the $(k-1)$th  jump, {\it i.e.},  in the beginning of 
the layer $k$  and before the jump, {\it i.e.} at the end of the layer $k - 1$.   
Finally, $D_k$ is the adiabatic evolution matrix of eigenstates in the layer $k$: 
\be
D_k = {\rm diag} (e^{- i \phi_k^m/2},  e^{i \phi_k^m/2}), 
\label{eq:diagd}
\ee
where $\phi_k$ is the adiabatic phase acquired in the layer $k$
\begin{equation}
\phi_k^m (E) \equiv \int_{x_{k-1}}^{x_k} \! \! d x \ \Delta_{21}^m(x). 
\label{phasexl}
\end{equation}
The diagonal character of $D_k$ reflects the adiabaticity of the neutrino propagation within the layers.

The change of mixing angle in the jump $j$ (\ref{eq:changeangle}) can be expressed in terms of change of 
the potential $\Delta V_j$  in the layer $j$  as 
\be
\sin \Delta \theta_j  \approx 
\Delta \theta_j \approx 
c_{13}^2 \sin 2 \theta_{12} \frac{E}{ \Delta m_{21}^2} \Delta V_j
\label{eq:mixchange}
\ee 
in lowest order in $\epsilon_m$.

The probability of the $\nu_1 \rightarrow \nu_e$ transition equals  
\be
P_{1e} = c_{13}^2|S_{e 1}|^2,   
\ee
where the factor $c_{13}^2$ follows from projecting back to the flavor basis: 
$\nu' \rightarrow \nu_f$. 

Let us compute the probability in the lowest order in $\epsilon_m$. 
Since $\Delta \theta_{k-1} \sim \epsilon_m$, the matrix of the basis change 
can written in the lowest order  as 
\be
U_{k, k-1} \approx I -  i \sigma_2 \sin \Delta\theta_{k-1},  
\label{eq:b-k-1}
\ee
where $\sigma_2$ is the Pauli matrix. Inserting this expression 
into (\ref{eq:ampl}) and keeping  only terms up to order $\epsilon$, 
we obtain 
\be
S = U_n^m (\theta^m_{12, n}) \left[D(\phi_{tot}) - i\sum _{j = 0}^{n-1} 
\sin \Delta\theta_{j} D(\phi_{j}^{a}) \sigma_2 D(\phi_{j}^{b})
\right],  
\label{eq:ampl1}
\ee
where we introduced summation over the jumps. 
Here $D$ are diagonal matrices of the form (\ref{eq:diagd}) with 
the total phase acquired in the Earth: 
\be
\phi_{tot} =  \sum_{k = 1}^n  \phi_{k}, 
\ee
and with the total phases acquired before and after jump $j$ respectively: 

\be
\phi_{j}^{b} =  \sum_{k = 1}^{j} \phi_{k}, ~~~~~\phi_{j}^{a} =  
\sum _{k = j + 1}^{n} \phi_{k}. 
\ee
Using (\ref{eq:ampl1}) we find explicitly the $e1$- element: 
\be
|S_{e1}| = \left| \cos \theta^{m,f}_{12, n}  + \sin \theta_{12, n}^{mf} 
\sum _{j = 0}^{n - 1} \sin \Delta \theta_j~ e^{- i\phi_j^a} \right|, 
\label{eq:ampl2}
\ee
where we have taken into account that $\phi_{tot} = \phi_{j}^{b} + \phi_{j}^{a}$. 
The total amplitude (\ref{eq:ampl2}) can be viewed as a superposition of waves emanating at  
the jumps. The amplitudes of the waves are determined by the sizes and signs of jumps, so that  
and the sign is positive (negative) if the potential increases (decreases) on the way 
of neutrinos.  

Finally,  the probability equals 
\be
P_{1e} =  c_{13}^2 \left[(\cos \theta_{12, n}^{m,f})^2  + \sin 2\theta_{12, n}^{m,f}
\sum _{j = 0}^{n - 1}\sin \Delta\theta_j \cos \phi_j^{a}\right].  
\label{eq:prob2}
\ee
According to  (\ref{eq:ampl2}) and (\ref{eq:prob2}), the probability is given 
by the zero order term and the sum of the contributions of the density jumps. 
The contribution from the individual jumps is  given 
by  the sine of change of the mixing angle in a jump (which is $\sim \epsilon_m$),   
and by the phase factor with the total phase acquired over the distance from 
a given jump to a detector (see also \cite{deHolanda:2004fd}). šlŠ

We will use these expressions for interpreting the
results of numerical computations in sect. III.

In Ref.  \cite{Ioannisian:2004jk},  \cite{Ioannisian:2004vv} the probability  
$P_{1e}$ has been obtained in more general integral form 
which describes both jumps effects and adiabatic propagation: 
 \begin{equation}
  \label{p1} 
P_{1e} =
       c_{13}^2\cos^2 \theta_{12} - {1 \over 2} \sin^2 2 \theta_{12}
     c_{13}^4 \int_{0}^{L} \! \! \! \! dx \  V_e(x) \sin \phi^m_{x \to L} ,
\end{equation}
where
\begin{equation}
\phi_{x \to L}^m (E) \equiv \int_{x}^{L} \! \! d x \ \Delta_{21}^m(x),
\label{phasexl}
\end{equation}
is the adiabatic phase acquired from a given point of trajectory $x$ to a detector 
at $L$.  $L = 2 R_E \cos \eta$ is the total length of trajectory, and $\eta$ is its nadir angle. 
This form is useful for derivation of  the 
attenuation effect (see below).

For the potential $V_e(x)$ with jumps  the integration 
in (\ref{p1}) can be performed explicitly which reproduces 
the result  (\ref{eq:prob2}).   
The phases (\ref{phasexl})  $\phi_{x_j \to L}^m$,  where $x_j$   is the coordinate of 
$j$-th jump,   coincide with
$\phi_{j}^{a}$.

The solar mixing parameter  $\cos 2 \bar{\theta}^\odot_{12}$ can be expressed in terms 
of $P_D$ using eq. (\ref{PD}). 
Then the difference of probabilities  (\ref{main3}) becomes  
\begin{eqnarray}
\Delta P (E, L) & = &   c_{13}^2 
(0.5 c_{13}^4 + s_{13}^4 - P_D) 
\nonumber
\\
& \times &  {\sin^2 2 \theta_{12} \over \cos 2 \theta_{12}}
 \int_{0}^{L} \!  dx \ V_e(x) \sin \phi^m_{x \to L}.  
 \label{DeltaP}
  \end{eqnarray} \\

A key element for understanding oscillations in the Earth is 
the {\it attenuation effect} \cite{Ioannisian:2004jk} which is 
a consequence  of  integrating $\Delta P$  with the neutrino energy 
reconstruction function $g(E_r, E)$
over the neutrino energy: 
\begin{equation}
\Delta \bar{P}(E_r) = \int dE g(E_r, E)  \Delta P(E) \ .
\label{avv}
\end{equation}
$E_r$ is the reconstructed energy of the neutrino. 
The function $g(E_r, E)$ is determined  by the following factors: 
(i) energy resolution of the detector, (ii) kinematics of reaction, (iii) 
energy spectrum of produced neutrinos. To see the effect of attenuation, 
we insert (\ref{DeltaP}) into  (\ref{avv}) and have 
\begin{eqnarray}
\Delta \bar{P}  & = & c_{13}^2  \left(0.5 c_{13}^4 + s_{13}^4 - P_D \right)
\nonumber
\\
& \times &  
{\sin^2 2 \theta_{12} \over \cos 2 \theta_{12}}
\int_{0}^{L} \!   dx \ V(x) { F(L-x)} \sin \phi^m_{x \to L},
\label{eq:attenuation}
\end{eqnarray}
which defines an attenuation factor $F(d)$ 
\cite{Ioannisian:2004jk}, with $d \equiv  L - x$ 
being the distance from the location of interest to a  detector. 
In (\ref{eq:attenuation}) the expression 
in parantheses has been  put out of the integral
because it depends only weakly on energy.
For the Gaussian energy resolution function 
\begin{equation}
 g(E_r, E)={1 \over \sigma_E \sqrt{2 \pi}}
e^{-{(E_r -E)^2 \over 2 \sigma_E^2}}
\label{eq:gaussian}
\end{equation}
we obtain 
\begin{equation}
F(d)\simeq e^{-2\left({ d \over \lambda_{att} }\right)^2},  
\nonumber
\end{equation}
where 
\begin{equation}
\lambda_{att} \equiv l_\nu \frac{E}{\pi \sigma_E}
\label{eq:attlength}
\end{equation}
is the attenuation length. For $d = \lambda_{att}$, the suppression
factor equals $F(d) = e^{-2} = 0.135$,  and according to   
(\ref{eq:attenuation}), the oscillatory effect  
of structures with $d > \lambda_{att}$  is strongly suppressed. 
As follows from (\ref{eq:attlength}),  the better the energy 
resolution of the detector, 
the more remote structures can be ``seen''. 

\begin{figure}[!]
\hspace{-1cm}
\includegraphics[width=0.4\textwidth, height=0.4\textwidth]{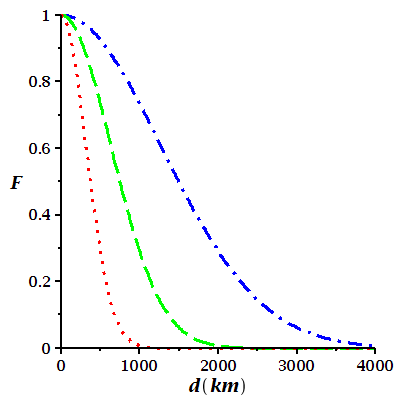}
\caption[�]{The attenuation factor $F$ as function 
of distance from a detector, 
$d$,  for $E = 11$ MeV and different values of the 
energy resolution $\sigma_E$:  0.5 MeV (blue line), 
1 MeV (green line), 2 MeV (red line). We take 
$\Delta m_{21}^2 = 7.5 \cdot 10^{-5} $ eV$^2$.
\label{fig:atten}
}
\end{figure}

In fig. \ref{fig:atten} we show the attenuation factor 
for different values of energy resolution. 
If $\sigma_E / E = 0.1$  and $l_\nu = 400$ km the attenuation 
length equals  1470 km 
and structures of the density profile at $d > 1470$ km can not be seen. 
For  $\sigma_E / E = 0.2$ the structures with  $d > 750 $ km 
are strongly attenuated.   

\section{III \ The relative excess of night events}

The relative excess of the night events (\ref{eq:excess})
as function of the nadir angle and reconstructed 
neutrino energy can be written as
\begin{equation}
A_{ND}(E_r, \eta) ={\int dE g(E_r, E) \ \sigma_{CC}(E) 
f_{B}(E) \Delta P (E) \over \int dE g(E_r, E) \ 
\sigma_{CC}(E) f_{B}(E) P_{D}(E)} . 
\label{eq:d-nasym}
\end{equation}
Here $f_{B}(E)$ is the boron neutrino spectrum \cite{Bahcall:1996qv}. 
Notice that only  9.7\% of $^8$B neutrinos have energy
$E_\nu$ $ >$ 11 MeV  but the 
corresponding fraction of the detected events is 0.9.

We compute the oscillation probabilities  $P_{D}(E)$ and $\Delta P(E)$  
according to (\ref{probnue}),  (\ref{PD}) and (\ref{DeltaP})  using   
the spherically symmetric model of the Earth with the eight layers parametrization 
of the PREM density profile \cite{Dziewonski:1981xy}.   
It  has 6 density jumps in the mantle $J_i^m$  
and 2 density jumps  $J^c_j$ in the core. The parameters of the jumps 
(depth, size, nadir angle of the trajectory 
which is tangential to the jump) are given in TABLE I.  
Trajectories with $\eta < 0.58$  cross the core of the Earth.  
The change of the solar neutrino flux due to the
eccentricity of the Earth orbit ($\pm 3.34 \%$) is taken into account.

For the energy resolution function $g(E_r, E)$ 
we use the Gaussian form (\ref{eq:gaussian})  
with different values of the width  $\sigma_E$. 

\begin{center}
\begin{table}
    \begin{tabular}{ |l|r|r|r|r|r| }
    \hline
Jump & depth($km$)& $\rho_- $(${g \over cm^3}$) & $\rho_+$(${g \over cm^3}$)  
& $2{\rho_+ -\rho_- \over \rho_+ + \rho_-}$ 
& $\eta_i$
     \\ \hline
     J$_0^m$ & 0 & 0  & 2.60 & & $\pi /2$
     \\ \hline
     J$^m_1$ & 15 & 2.60 & 2.90 & 0.11 & 1.50
     \\ \hline
     J$^m_2$ & 25 & 2.90 & 3.38 & 0.15 & 1.48
    \\ \hline 
    J$^m_3$ & 220 & 3.36 & 3.44 & 0.02 & 1.31
    \\  \hline
    J$^m_4$ & 400 & 3.54 & 3.72 & 0.05 & 1.21
    \\ \hline
    J$^m_5$ & 670 & 3.99 & 4.38 & 0.09 & 1.11
    \\ \hline
    J$^c_1$ & 2891 & 5.57 & 9.90 &  0.56 & 0.577
    \\ \hline
    J$^c_2$ & 5150 & 12.17 & 12.76 & 0.05 &  0.193
    \\ \hline
    \end{tabular}
    \caption[...]{
Parameters of the density jumps used in our computations:  
the depth from the surface, density before a jump,  
$\rho_-$, the density after the jump $\rho_+$, relative size of the jump, 
nadir angle at which the neutrino trajectory 
touches the surface of a jump.   
The eight jumps correspond to the surface of the Earth,  
the outer crust, the inner crust, lid, low velocity zone, 
transition zone, low mantle, outer core, inner core. 
}
\end{table}    
\end{center}

For cross-sections we take a generic form for interaction with 
nuclei: 
\begin{equation}
\sigma_{CC} (E) = A {p_e E_e},
\label{eq:crossect}
\end{equation}
where $A$ is a normalization factor (irrelevant for the relative excess), 
$E_e = E_\nu - \Delta M$, $p_e$ is the electron momentum  and $\Delta M$ is the threshold  
of reaction. In computations for DUNE we use $\Delta M = 5.8$ MeV (see below).

\begin{figure*}[!]
\begin{center}
\includegraphics[width=0.49\textwidth]{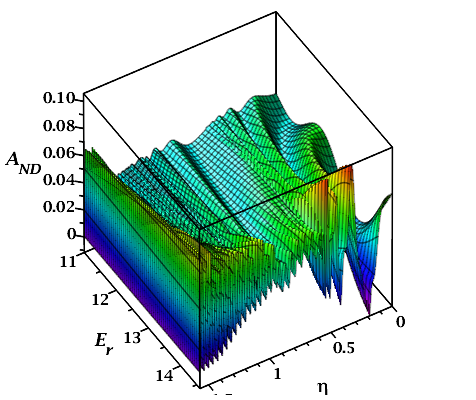} \\
\includegraphics[width=0.49\textwidth]{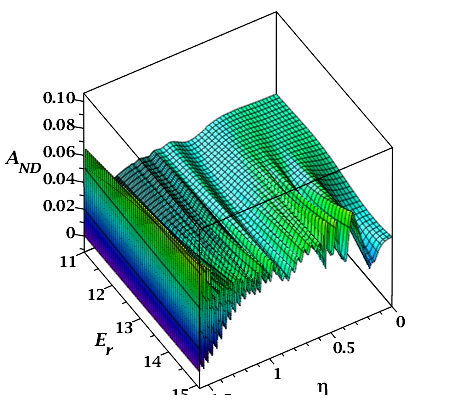}\\
\includegraphics[width=0.49\textwidth]{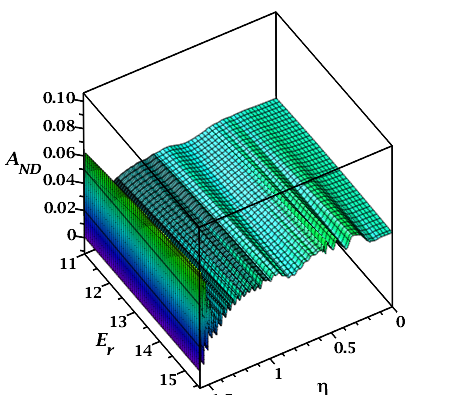}  %
\end{center}
\caption[...]{The relative excess of the night events 
produced by boron neutrinos 
as function of the nadir angle 
and reconstructed neutrino energy for different  
energy resolutions:  
$\sigma_E = 0.5$ MeV (upper panel),  1 MeV (middle panel) and 2 MeV 
(bottom panel).  The Gaussian form of the reconstruction function is used.   
\label{fig:dn2dim}
}
\end{figure*}

\begin{figure*}[!]
\begin{center}
\includegraphics[width=0.8\textwidth]{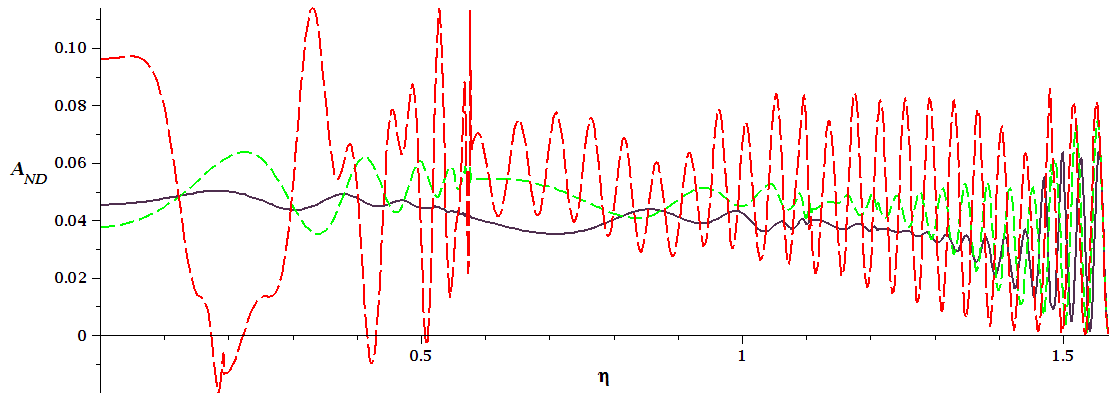}\\
\includegraphics[width=0.8\textwidth]{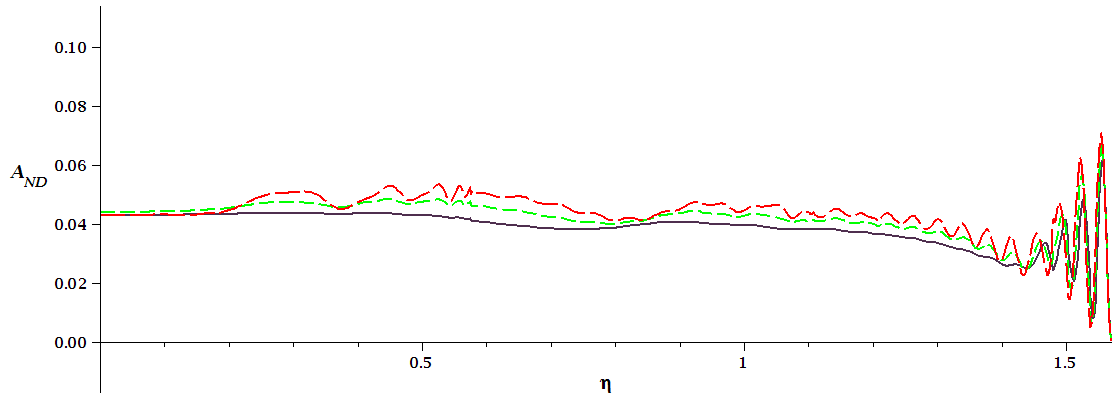}
\end{center}
\caption[...]{The relative excess of night events  
as function of the nadir angle for different 
values of the reconstructed energy: 
$E_r$ = 11 MeV (violet solid line),  
13 MeV (green dash line), and 15 MeV (red long-dash line).
The upper (lower) panel corresponds to  the energy resolution 
$\sigma_E = 0.5$ MeV ($\sigma_E = 2$ MeV). 
\label{fig:zenith-det}
}
\end{figure*}

Notice that at the energies of Boron neutrinos 
$\cos 2 \bar{\theta}^\odot_{12} < 0 $ and therefore 
according to (\ref{main3}) the regeneration (increase) of the $\nu_e$ flux 
$\Delta P > 0$ corresponds to suppression of $P_{1e}$: 
due to the Earth matter effect 
$P_{1e} <  P_{1e}^0$.

In Fig. \ref{fig:dn2dim}  we show the relative 
excess of the night events,  
$A_{ND}$ (\ref{eq:d-nasym}), 
as function of the nadir angle and the reconstructed neutrino  energy
for different values of the energy resolution
$\sigma_E$. The excess increases with energy.   
Due  to attenuation, remote  structures  are not seen
for poor resolution (bottom panel). In particular, 
the core of the Earth is not noticeable
with $\sigma_E = 2$ MeV, and the dependence on $\eta$ is as if the core would be absent.
With increase of energy, a small oscillatory effect 
appears due the core  at $\eta < 0.58$ since the attenuation is determined
by the relative resolution ($\sigma_E/E$). 
Details of the $\eta-$ dependence for different energies can be seen 
in Fig. \ref{fig:zenith-det}. 
Notice that the core of the Earth is clearly visible for 
$E = 15$ MeV and $\sigma_E = 0.5$ MeV.

\begin{figure}[h]
\includegraphics[width=0.45\textwidth]{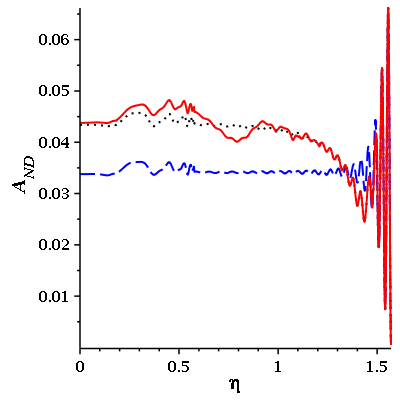}
\caption[�]{The  
relative excess of the night events 
integrated over  $E_r > $11 MeV as function 
of the nadir angle of the neutrino trajectory 
for different density profiles: 
PREM profile (red solid line); profile without     
density jumps at 400 and 660 km (black dot line) and with  density 
jumps at outer and inner cores only (blue dash line).
\label{fig:zen-indj}}
\end{figure}

Fig. \ref{fig:zen-indj} (red line)
shows the $\eta$ distribution integrated over the energy above 11 MeV.  
This integration is equivalent to  an energy resolution function  
of  a box-like form  with  the average energy 12.5 MeV and 
energy resolution $\sigma_E = 1.5$ MeV, {\it i.e.},  $\sigma_E/E = 0.12$. 
The corresponding attenuation length equals $\lambda_{att} = 1200$ km. 
The form of the distribution is rather generic.  
Different computations (including those for  SuperKamiokande) 
give rather similar pattern which can be understood in the following way. 

Because of the linearity of the problem, the integration over the energy 
and the integration of the evolution equation  can be permuted.  
That is, one can first integrate over energy obtaining 
the attenuation and then consider  the 
flavor evolution, or first,  compute the flavor evolution 
and then perform the energy integration.

Without density jumps ("no-jump" case) the $\eta-$distribution 
would have a  regular oscillatory pattern  
with a constant averaged value of ${A}_{ND}$ determined by the surface density,  
and the depth of oscillations which decreases 
with decreasing $\eta$ (the attenuation effects are stronger for longer trajectories). 
In the realistic case this (regular oscillatory pattern)
occurs only for $\eta  > 1.50$ 
when the neutrino crosses  a single outer layer. 
The period of the oscillatory curve in $\eta$, $\Delta \eta$, 
can be estimated from the condition 
$\Delta L = 2 R_E \sin \eta  \Delta \eta = l_m$,  which gives 
$$
\Delta \eta = \frac{l_m}{2R_R \sin \eta}.
$$
For $\eta \rightarrow 0$ 
(approaching the core) the 
period increases.  
The blue dashed line in Fig.~\ref{fig:zen-indj} illustrates 
such a  behavior down to 
$\eta = 0.58$ below which small perturbations appear due 
to the core effect.  

The jumps break adiabaticity and modify the above picture.  
The deviations start at $\eta =  1.50$, the 
nadir angle of the neutrino trajectory 
which touches the surface of the first jump. 
The length of the  trajectory is  $L_1 = 875$ km,  
{\it i.e.},  approximately  $ \approx 2 l_m$.

Attenuation leads to  hierarchy of these  jump effects. 
The jumps $J_1^m$ and $J_2^m$, closest to a detector,
produce the strongest effects (see dotted line in Fig.~\ref{fig:zen-indj}).  
In turn, jumps  $J_4^m$ and $J_5^m$ weakly perturb 
the picture produced by $J_1^m$ and $J_2^m$ 
at $\eta < 1.21$ (the density change of $J_3^m$ is too small). 
Finally,  the remote core jumps $J_1^c$ and $J_2^c$ modifiy   
the previous picture  at $\eta < 0.58$ even more weakly. 
Let us consider these modifications in order.

As noticed before, the strongest modifications of the ``no-jump'' picture   
is  produced by the jumps $J_1^m$ and $J_2^m$ at $\eta = 1.4822$ 
and  $1.502$ respectively. They suppress the  oscillatory behavior 
at $\eta >  1.45$, produce a dip at 
$\eta = 1.3 - 1.5$,   and lead to  smooth increase of the excess 
(above no-jump case) with decrease of $\eta$. 
(See dotted line in Fig. \ref{fig:zen-indj}). 

For $\eta >  1.45$ the length of the trajectory 
is less than the attenuation length; therefore 
all jumps should be taken into account. 
For $1.502 > \eta >  1.482$ neutrinos cross three layers and
at $\eta < 1.482$ -- five layers of matter.
Furthermore, for the effective energy $E = 12.5$ MeV the oscillation length
equals 430 km which is comparable to the lengths of sections of the trajectory in the layers.  
This leads to a {\it parametric suppression} of 
oscillations in addition to averaging:
the jumps suppress  the third and fourth oscillation  maxima 
in accordance with Fig. \ref{fig:zen-indj} (red line).  
For parametric effects, in general, see \cite{paramet}, \cite{Krastev:1989ix}.

To illustrate how it works, let us consider the third maximum of $A_{ND}$ (red line in Fig.\ref{fig:zen-indj}) at  $\eta = 1.489$. 
For this  $\eta$  the neutrino trajectory crosses 
three density jumps: $j = 0$, $J_0^m$ at the surface, $j = 1$, $J_1^m (far)$, and  
$j = 2$, $J_1^m (near)$. This corresponds to crossing three layers:  
the outer layer twice (at the beginning and the end) 
and the second one in between. 
The total length of trajectory  is $L \approx 2.5 l_m$, 
the lengths of the trajectory sections in individual layers 
equal 238 km, 562 km  and 238 km or approximately  
$0.5 l_m$, $1.5 l_m$ and $0.5 l_m$. 
Therefore the phases acquired from the three jumps to a detector 
equal $\phi_0^m = 5\pi$, $\phi_1^m  = 4\pi$ and $ \phi_2^m  = \pi$.  
Inserting these numbers into 
the expression for probability (\ref{eq:prob2}) we obtain
\begin{eqnarray}
P_{1e} & = & c_{13}^2 \left[(\cos \theta_{12, n}^m)^2  + \right.
\nonumber\\
&& \hskip-1cm \sin 2\theta_{12, n}^m (-\sin \Delta\theta_0  
\left.  + \sin \Delta\theta_1  + |\sin \Delta\theta_2|)  \right].
\label{eq:prob3}
\end{eqnarray}
Here the first term in the parenthesis in the second line is the  contribution 
from the jump $j = 0$. Since 
$\phi_0^m = 5\pi$ and density increases on the way of the neutrinos
($\sin \Delta\theta_0 > 0$), the 
contribution is negative, thus leading to the positive contribution to the night events excess.
The second term, from jump $j = 1$,  is positive: now $\phi_0^m = 4\pi$ and 
the density increases ($\sin \Delta\theta_1 > 0$).   
The third term (from $j = 2$) is again positive,  since 
$\phi_0^m = \pi$ and density in this jump decreases, so that 
$\sin \Delta\theta_2 < 0$.  Thus, internal jumps suppress the excess ($\nu_e$-regeneration). 
In other words, the waves ``emitted'' from the surface jump 
and the two internal jumps interfere destructively. 

\begin{figure}[h]
\includegraphics[width=0.4\textwidth]{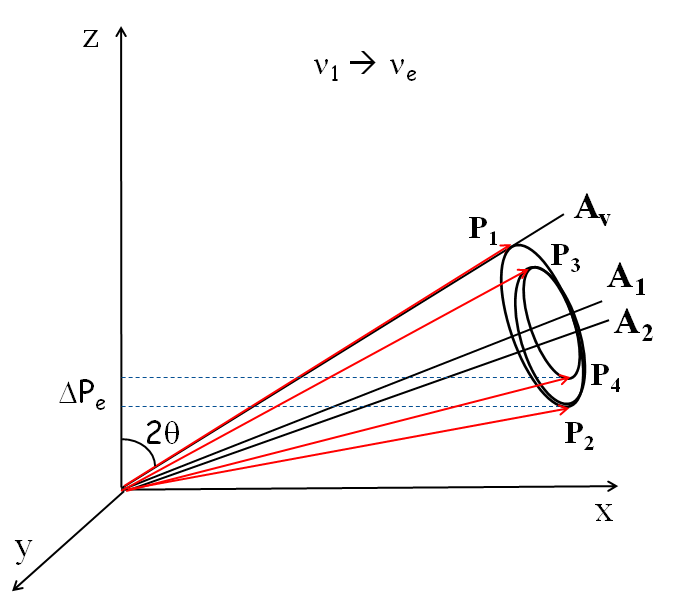}
\includegraphics[width=0.4\textwidth]{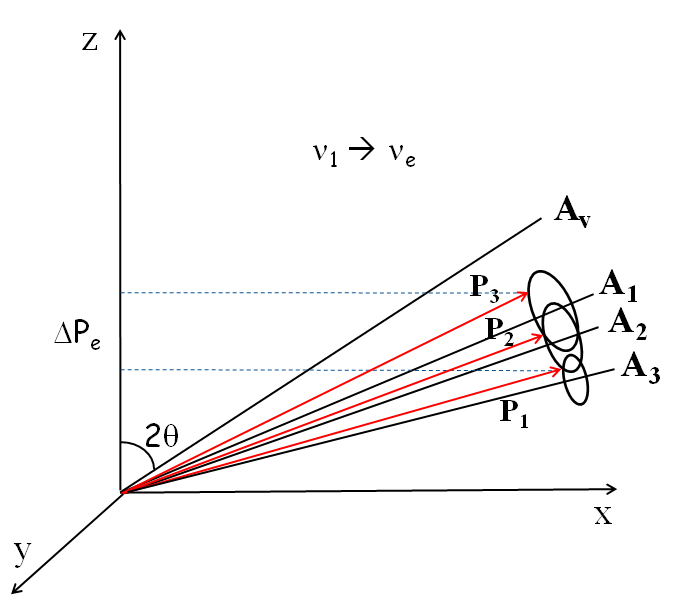}
\caption[�]{Graphic representation of the neutrino oscillations 
in the Earth. Shown are the positions   
of neutrino polarization vector ${\bf P}_i$  at the borders of 
different layers.  
${\bf A}_i$ is  the  precession axis in the layer $i$.    
The upper panel: the parametric suppression of oscillations for $\eta = 1.489$;  
the bottom panel: formation of the dip. See further details in the text. 
\label{fig:graphic}}
\end{figure}

The effect can be visualized by  an analogy with the 
electron spin precession in the magnetic field \cite{ms4}, 
\cite{Bouchez:1986kb}, \cite{Krastev:1989ix}. 
In this representation the neutrino state is described by a 'polarization vector' ${\bf P}$ in flavor space
(${\bf x}, {\bf y}, {\bf  z}$) (see Fig.~\ref{fig:graphic}) whose length is $|{\bf P}| = 1/2$.
The probability to find $\nu_e$ in this state is given by the projection of ${\bf P}$ on the axis ${\bf z}$ (the flavor axis) as  $P_e = 0.5 + P_z$.
In the  layer with  a given matter density $\rho$, ${\bf P}$ precesses around the axis ${\bf A}$, the direction of eigenstates in this layer. The axis ${\bf A}$ lies in the plane (${\bf x} - {\bf  z}$). The angle between ${\bf A}$ and ${\bf  z}$ is twice the flavor mixing angle in matter, $2\theta_{12}^m$. In vacuum, the angle between the  axis of eigenstates (mass eigenstates) ${\bf A_v}$  and ${\bf  z}$ is $2\theta_{12}$. The angle of precession coincides with the oscillation phase.  At the borders between layers the mixing angle in matter, and correspondingly,  the direction of the axis of eigenstates sharply change.

According to  Fig.~\ref{fig:graphic}, upper panel) a 
neutrino $\nu_1$ entering the Earth is described by the polarization vector ${\bf P}_1$.  
In the first layer the vector precesses  around ${\bf A}_1$ 
by half a period.   
So, at the border with the second layer it  
reaches the position ${\bf P}_2$. In the second layer the precession proceeds 
around axis ${\bf A}_2$ (whose direction with respect to axis ${\bf z}$ 
is determined by the corresponding mixing 
angle in matter). The neutrino vector precesses here 
by 1.5 periods and therefore it enters the third layer 
in the state  ${\bf P}_3$. Since the  layer 3  has the same properties as the 
first layer neutrino vector precesses there again by half a period 
around  ${\bf A}_1$  and reaches a  detector in the state ${\bf P}_4$.   
(Notice that after crossing  each layer the opening angle of precession 
cone systematically decreases.)  
In the absence of the internal jumps neutrino would be in position ${\bf P}_2$. 
The projection of difference $({\bf P}_4 - {\bf P}_2)$ onto the flavor axis ${\bf z}$ is 
positive, thus leading to suppression of the night excess.

\begin{figure}[h]
\includegraphics[width=0.45\textwidth]{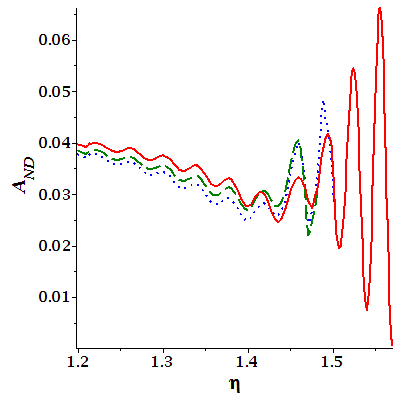}
\caption[�]{The relative excess of night events 
integrated over  $E_r > $11 MeV as function 
of the nadir angle of the neutrino trajectory 
for changed parameters for the density jumps. The
red line is the conventional one with jumps at 15 km and 25 km; 
blue-dotted line is for jumps  at  $h_1 = 20.5$ km and $h_2 = 30$ km; 
green-dashed  line is  for jumps at $h_1 = 15$ , $h_2 = 30$. 
Parametric enhancement of oscillations is seen in the 3rd and 4th periods. 
\label{fig:zen-int1}}
\end{figure}

This picture can be modified by local density perturbations near the detector.   
A variation of the depth of the jumps (distance from the surface) can further modify 
the $\eta$ dependence  
leading, e.g.,  to a parametric enhancement (rather than suppression) 
of oscillations (see Fig. \ref{fig:zen-int1}).

The dip at $\eta \sim 1.4$ is the interplay of several  factors: 

(i) In the region of $\eta = (1.20 - 1.45)$ 
after entering the Earth neutrinos cross 4 jumps in the following order: 
$J_1^m (far)$,  $J_2^m(far)$, $J_2^m(near)$,  $J_1^m(near)$. 
Here ``far'' and  ``near'' determine position of a jump  with respect to a detector.   
At $\eta \sim 1.4$ the length of the trajectory,  $L = 2160$ km,  
is substantially bigger than the attenuation length. 
Therefore the remote jumps  $J_1^m(far)$  and  $J_2^m(far)$ 
(close to the point where neutrinos entered the Earth) 
are not seen because they are attenuated. Consequently, the effect at a  detector  
is determined by oscillations 
in the third layer and the jumps closest to a detector
$J_2^m(near)$  and  $J_1^m(near)$.

(ii) Oscillations in the third layer (between $J_2^m(far)$ and  $J_2^m(near)$) 
are strongly averaged, so the corresponding oscillation amplitude is effectively small.
Thus, the  opening angle  of the precession cone  with respect to axis ${\bf A}_3$ 
in Fig.~\ref{fig:graphic}b)  is small. 
Furthermore, this angle  is smaller than the angle  
between axes $A_3$ and  $A_2$.

(iii) The length of the trajectory in the outer layers is smaller than $l_\nu/2$
(for $\eta > 1.4$  the length  is even smaller than $l_\nu/4$).
In this case and also because of (ii) jumps $J_2^m (near)$ and $J_1^m(near)$
with decreasing densities  
systematically pull the neutrino 
vector up -- to the initial state, {\it i.e.} 
suppressing transition (Fig.~\ref{fig:graphic}, bottom panel).
After oscillations in the layer 3 around axis ${\bf A}_3$ 
the neutrino vector enters the layer 4  
in the state ${\bf P}_1$. 
(Because of smallness of radius of precession 
similar result will be obtained for any position of ${\bf P}_1$ on the precession cone.) 
It precesses around ${\bf A}_2$ ($\equiv {\bf A}_4$) 
by about 1/4 of the period, from 
the state ${\bf P}_1$ to ${\bf P}_2$. 
In the state ${\bf P}_2$ the neutrino enters the layer 5 and precesses around 
${\bf A}_1$ (which is the same as ${\bf A}_5$) by less than 1/4 of the period. It reaches a detector 
in the state ${\bf P}_3$.  The projection  of the difference 
$({\bf P}_4 - {\bf P}_1)$ onto the flavor axis is positive 
implying suppression of the excess. 

For $\eta < 1.2$ the lengths of trajectories in the first two layers 
become much smaller than the oscillation 
length ($l_i/l_m < 0.1$) and in the first approximation 
the oscillation effect in these layers can be neglected. 
In this case one can consider oscillations in the third layer only  
with initial density as  at the border 
of this layer, that is,  $\rho_3 = 3.4~ g/cm^3$. 
Propagation in the layer 3 is adiabatic 
and therefore the average oscillation 
effect is determined by its surface density. 
Consequently,  the average oscillation 
effect here will be bigger than the average effect at the surface
by factor $\rho_3/ \rho_1$. This determines the asymptotics   
of $A_{ND}$ at small $\eta$:
\be
{A}_{ND} (\eta = 0) = \frac{\rho_3}{\rho_1} A_{DN} (\eta = 1.57)  
\ee
and $\rho_3/\rho_1  = 1.31$. 
So, for $\eta$ below the  dip the averaged excess  will increase 
approaching  $4.7 \%$ as compared to  
$3.55\%$ at the surface, in agreement with results 
of computations (Fig.~\ref{fig:zen-indj}).   

The pattern  produced by $J_1^m$ and $J_2^m$ is further perturbed by the
deeper  jumps $J_4^m$ and $J_5^m$. They cause a
modulations of the distribution.   
The jump  $J_4^m$ has smaller size than $J_5^m$,
but it is closer to the  detector and  
therefore  the two modulations have comparable size. Since $J_4^m$  is closer 
than  $J_5^m$ to a detector, its modulations have 
a larger period than those of $J_5^m$ for the same range 
of $\eta$. For $\eta \sim 0.8$ the modulations 
are in phase leading to significant dip. 
At larger $\eta$ they are out of phase, reducing the modulations. 

The contribution of the core jump $J_1^c$, in spite of its large size,  
is strongly attenuated, resulting in 
even smaller perturbation on the top of those 
generated by mantle jumps at $\eta < 0.58$.  

These qualitative consideration show clearly the sensitive connection 
between features of the Earth density profile
and the $\eta$ distribution.

\section{IV \ Physics reach of DUNE}

We will assume detection based on the neutrino - nuclei 
interactions which have several advantages: 
(i) they have good neutrino energy resolution/reconstruction; 
(ii) the cross-section is much larger than the $\nu e- $ scattering cross-section, 
(iii) the damping factor due to contribution 
of $\nu_\mu$ and $\nu_\tau$ is absent. 
Correspondingly the asymmetry is enhanced by factor 
$[1 + \kappa /(P_D(1 - \kappa))] \approx 1.6$,  
where $\kappa \equiv  \sigma_{NC}/\sigma_{CC}$ 
is the ratio of neutral to charged current
cross-sections of  scattering on electrons.

To illustrate the potential of our method, we consider a future DUNE experiment \cite{Acciarri:2015uup}.  
At DUNE solar neutrinos are detected by the CC process
\begin{equation}
\label{QE}
\nu_e + ^{40}{\rm Ar} \to ^{40}{\rm K} + e^- . 
\end{equation}
Since $^{40}$Ar has spin 0 and the ground state of $^{40}$K has 
spin 4, the transition to the ground state 
is highly suppressed. 
The transition(s) via intermediate excited states of $^{40}$K 
(Fermi or Gamov-Teller transitions) with further 
emission of photons are more probable, although the rates of these transitions are not measured yet.

\begin{figure}[!]
\hspace{-1cm}
\includegraphics[width=0.4\textwidth, height=0.35\textwidth]{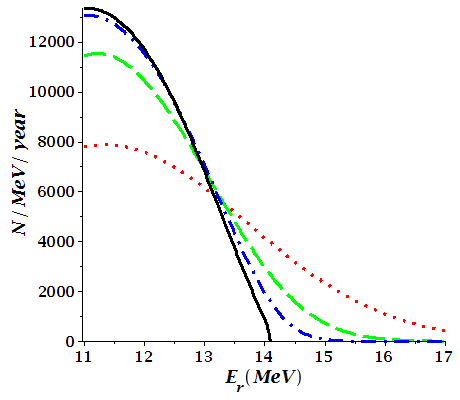}
\caption[É]{
The energy ($E_R$) distribution of the annually detected events at DUNE   
for different energy resolutions $\sigma_E$. The solid (black) line represents  
perfect resolution,  $\sigma_E =  0$, the other lines correspond to  
$\sigma_E =  0.5$ MeV -   
dash-dotted (blue) line, $\sigma_E =  1$ MeV dashed (green) line, 
$\sigma_E =  2$ MeV dotted (red) line.
The distributions are normalized to 
annual number of events 27000 at $E_r >$ 11 MeV. 
\label{Fig7}
}
\end{figure}

We approximate the cross-section as in (\ref{eq:crossect}) with 
$A \simeq 2 \cdot 10^{-43} {\rm cm}^2 {\rm MeV}^{- 2}$.   
$\Delta M  = 5.8$ MeV is the mass  difference 
between the Fermi exited $^{40}$K state and  $^{40}$Ar. 
We consider neutrinos with energy $E_\nu > $11 MeV,  
since an  electron, to be detected,  should have energy above 5 MeV 
(\ref{QE}).  
  
We find that about 27000  $\nu_e$ events will be detected annually 
for  $E_\nu > $ 11 MeV in a 40 kt liquid argon detector due 
to the reaction (\ref{QE}).
The energy distribution of these events is shown in Fig. \ref{Fig7}.  
Correspondingly, in 5 years 135000 events will be recorded.

With this statistics the following studies can be performed.

\begin{figure}[h]
\hspace{-1cm}
\includegraphics[width=0.4\textwidth, height=0.3\textwidth]{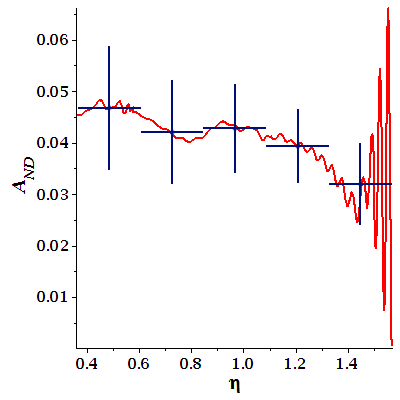}\\
\hskip-0.8cm
\includegraphics[width=0.4\textwidth, height=0.3\textwidth]{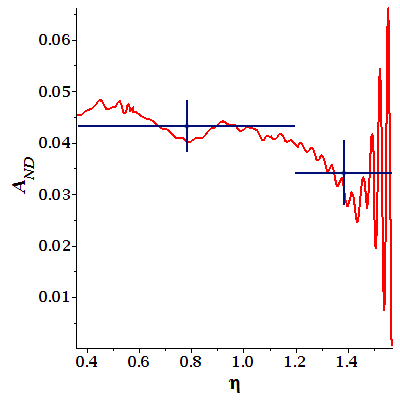}
\caption[...]{
Scanning of the Earth density profile. The crosses denote the average values of the relative night event excess
over the
$\eta$-interval given by the horizontal line; the vertical lines give the expected accuracy after 
5 years of data taking. 
\label{scan}
}
\end{figure}

\begin{figure}[h]
\hspace{-1cm}
\includegraphics[width=0.4\textwidth, height=0.3\textwidth]{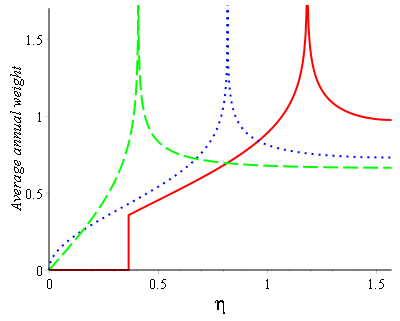}
\caption[...]{
Annual exposure time as function of nadir angle for different positions of 
a detector: 
Homestake mine (red solid line), tropics (dotted blue line),  
equator (dashed green line). 
\label{exposure}
}
\end{figure}

1). The integral relative excess (Day-Night asymmetry) 
can be measured  with high accuracy.  
After 5 years of data taking  statistical error will be $0.6 \%$. 
Therefore the average value $A_{ND} =  4\%$ will be measured with 
$15\%$ accuracy, and its deviation from zero will be established 
at more than a $6\sigma$ level. If $\Delta m^2_{21} = 
5 \cdot 10^{-5}$ eV$^2$,  the average excess equals  
$6.5 \%$ and accuracy of its measurement will be $10\%$  
($11\sigma$ from 0).

The background is largely unknown, although it is expected 
that it  will have no day-night variations 
which could mimic or modify the true Earth matter effect.   
If  the signal to background ratio is $S:B =  1:3$,  
the error bars will be 2 times larger:   
$1.2\%$. Consequently, $A_{ND}$ 
will be measured with $30\%$ accuracy ($20\%$ for high values of $\Delta m^2_{21}$).

It is straightforward to scale these results with increase of exposure. 

2). Measurements of $\Delta m_{21}^2$ and search for new physics 
effects. The difference of $A_{ND}$ between high and low values of 
$\Delta m_{21}^2$ is $\Delta A_{ND} = 2.5\%$. After 5 year 
the difference can be established  at $4\sigma$ level. 
If DUNE will confirm the low (solar) value,  
whereas JUNO will obtain high (KamLAND) value,   
this will be further evidence of new physics effects, 
e.g., non-standard neutrino interactions.

3).  Seasonal variations of the flux due 
to the elliptic orbit of the Earth around the Sun (3.34 \% amplitude)  
will be established at $(5 - 6)\sigma$ level.

4). The measurement of the nadir angle distribution 
of the excess integrated over the energy can be performed with 
$\sim 30\%$ accuracy (see Fig. \ref{scan}, upper panel).
Here, the exposure function shown in  Fig.~\ref{exposure} is important. 
However, we expect the detection of various structures of the
Earth profile will be difficult. 
The hope is that at least the main feature --
the dip  and increase of the excess 
with decrease of $\eta$ produced by two closest to surface jumps 
will be established. 
In Fig. \ref{scan} (bottom panel) we divided the entire interval 
accessible to DUNE into two bins 
$\eta =  (0.365 -1.2)$ and $\eta = (1.2 - 1.57)$ whose fractions 
of events are  $60\%$ and $40\%$, respetively. 
This difference of the numbers of events in the two bins can be established at about 
$2\sigma$ level.

5). The dependence on energy can be explored; measuring the $\eta$-distributions in two, 
or even more, energy intervals
seems feasible. \\

Measuring the hep neutrino flux will be difficult.
Annually, about 130 such neutrinos are expected in the energy interval 
$14.1 < E_\nu <18.8$ MeV. Very good energy neutrino reconstruction 
and high statistics are needed to disentangle 
them from events produced by the boron neutrinos.

The outer core ($\eta < 0.58$ ) is "visible" at the Homestake 
site at about $9\%$ of night time.
However,  its observation would also require high energy resolution and 
high statistics. 

Another possible realization of  $\nu$ - nuclei detection  
is the ASDC -THEIA -  future advanced
scintillation detector concept 
which uses water based liquid scintillator (WbLS) \cite{Alonso:2014fwf}.
The  WbLS can be loaded by metallic ions, in particular $^7$Li,  which will allow to
detect the CC process  $\nu_e + ^7{\rm Li} \rightarrow e + ^7{\rm Be}$.

\section{V \ Conclusion}

1. In view of forthcoming and planned solar neutrino 
experiments with large mass detectors and good energy resolution 
we performed a detailed study of
oscillations of the $^8$B neutrinos in the matter of the Earth.
As a model density profile for the earth, we have taken 
the so called PREM model  \cite{Dziewonski:1981xy}
which approximates the earth by several shells of slowly varying densities.
For such a profile we can nicely represent the oscillation amplitude as 
superposition of the waves emanating from the density jumps 
between the shells.   
We have computed the relative excess of the night events 
(Day-Night asymmetry)
as function of the reconstructed neutrino energy and the nadir
angle. 
Also we have computed  the nadir angle distribution of events 
integrated over the energy above 11 MeV.

2. The observable distribution are strongly affected 
by two major effects: Attenuation (due to the finite energy resolution of detectors)
and a parametric suppression (or enhancement)
of oscillations in the multi-layer medium (due to the interplay of wave length and thickness of the layers).

3. The density jumps influence the distribution substantially. 
Due to attenuation which affects more contributions from 
the far away (from the detector) jumps, there is a 
hierarchy of perturbations determined by the closeness of jumps to a  detector. 
Therefore in a first approximation, the $\eta$ dependence of the 
excess is given  by the two  jumps nearest to the detector. 

The $\eta$-distribution has the following generic properties:

- regular oscillatory pattern for $\eta > 1.502$  (the longer trajectories)
with strongly decreased amplitude due to integration over energy.

- dip at $\eta = 1.4$ which is due to attenuation of remote jumps 
and effect of closest jumps where the density decreases when neutrino pass them.

- increase of the relative excess with decreasing 
$\eta$ below $1.4$ and approaching 
the constant value determined by the density at the borders of third layer. 

- for $\eta < 1.2$ this first order picture is modulated 
by smaller effects of two deeper jumps in the mantle.

- for $\eta < 0.58$ further perturbations of the above picture show up 
due to the core of the Earth.

4. We computed the relative  excess of events and its $\eta$ 
distribution at DUNE. 

After 5 years of data taking DUNE can establish 
the DN-asymmetry at the $6\sigma$ level.  
The low and high values of $\Delta m^2_{21}$ 
can be distinguished at $4\sigma$ level.

The first scanning of the Earth matter profile will be possible: 
The nadir  angle distribution can be measured with $30\%$ accuracy. 
This will allow to establish the dip in the distribution at $2\sigma$ 
level. 

Further developments of the experimental techniques 
are required to get information about sub-dominant 
structures of the  $\eta$ distribution produced by inner mantle 
jumps and the core. \\

\section*{References}



\begin{thebibliography}{99}


\bibitem{ms4} S. P. Mikheyev and A. Yu. Smirnov, Proc. of the 6th
  Moriond Workshop on massive Neutrinos in Astrophysics and Particle
  Physics, Tignes, Savoie, France Jan. 1986 (eds. O. Fackler and
  J. Tran Thanh Van) p. 355 (1986).


\bibitem{Carlson:1986ui}
  E.~D.~Carlson,
  Phys.\ Rev.\ D {\bf 34} (1986) 1454.


\bibitem{Cribier:1986ak}
  M.~Cribier, W.~Hampel, J.~Rich and D.~Vignaud,
  Phys.\ Lett.\ B {\bf 182} (1986) 89.

\bibitem{Bouchez:1986kb}
  J.~Bouchez, M.~Cribier, J.~Rich, M.~Spiro, D.~Vignaud and W.~Hampel,
  Z.\ Phys.\ C {\bf 32} (1986) 499.

\bibitem{Hiroi:1987pe}
  S.~Hiroi, H.~Sakuma, T.~Yanagida and M.~Yoshimura,
  Prog.\ Theor.\ Phys.\  {\bf 78} (1987) 1428.
\bibitem{Baltz:1986hn}
  A.~J.~Baltz and J.~Weneser,
  Phys.\ Rev.\ D {\bf 35} (1987) 528.

\bibitem{Dar:1986pj}
  A.~Dar, A.~Mann, Y.~Melina and D.~Zajfman,
  Phys.\ Rev.\ D {\bf 35} (1987) 3607.

\bibitem{ms87}
S. P. Mikheyev and A. Yu. Smirnov, Proc. of 7th Moriond Workshop on Search for 
New and Exotic Phenomena,  Les Arc, Savoie, France, 1987, edited by O. Fackler
and J. Tran Thanh Van
(Editions Frontieres, Gif-sur-Yvette, France, 1987) p. 405.


\bibitem{Baltz:1988sv}
  A.~J.~Baltz and J.~Weneser,
  Phys.\ Rev.\ D {\bf 37} (1988) 3364.


\bibitem{Renshaw:2013dzu}
  A.~Renshaw {\it et al.} [Super-Kamiokande Collaboration],
  Phys.\ Rev.\ Lett.\  {\bf 112} (2014) no.9,  091805
  [arXiv:1312.5176 [hep-ex]].
 
 
\bibitem{Abe:2016nxk} 
  K.~Abe {\it et al.} [Super-Kamiokande Collaboration],
  Phys.\ Rev.\ D {\bf 94}, no. 5, 052010 (2016)
  [arXiv:1606.07538 [hep-ex]].
  

\bibitem{Ahmad:2002ka}
  Q.~R.~Ahmad {\it et al.} [SNO Collaboration],
  Phys.\ Rev.\ Lett.\  {\bf 89} (2002) 011302
  [nucl-ex/0204009].

\bibitem{Aharmim:2005gt}
  B.~Aharmim {\it et al.} [SNO Collaboration],
  Phys.\ Rev.\ C {\bf 72} (2005) 055502
  [nucl-ex/0502021].



\bibitem{Maltoni:2015kca}
  M.~Maltoni and A.~Y.~Smirnov,
  Eur.\ Phys.\ J.\ A {\bf 52} (2016) no.4,  87
  [arXiv:1507.05287 [hep-ph]].

\bibitem{Gando:2010aa}
  A.~Gando {\it et al.} [KamLAND Collaboration],
  Phys.\ Rev.\ D {\bf 83} (2011) 052002
  [arXiv:1009.4771 [hep-ex]].

\bibitem{An:2015jdp} 
  F.~An {\it et al.} [JUNO Collaboration],
  J.\ Phys.\ G {\bf 43}, no. 3, 030401 (2016)
  [arXiv:1507.05613 [physics.ins-det]].

\bibitem{Dziewonski:1981xy}
  A.~M.~Dziewonski and D.~L.~Anderson,
  Phys.\ Earth Planet.\ Interiors {\bf 25} (1981) 297.

\bibitem{shearer}
P. M. Shearer, (2013) Upper Mantle Seismic Discontinuities,
in Earth's Deep Interior: 
Mineral Physics and Tomography From the Atomic to the Global Scale 
(eds S.-I. Karato, A. Forte, R. Liebermann, G. Masters and L. Stixrude), 
American Geophysical Union, Washington, D. C.. doi: 10.1029/GM117p0115. 

\bibitem{petersen1993}
N. Petersen, et al., Sharpness of the mantle discontinuities, 
Geophys. Res. Lett., 20, 859-862,1993.



\bibitem{Bezrukov:2016gmy}
  L.~Bezrukov and V.~Sinev,
  Phys.\ Part.\ Nucl.\  {\bf 47} (2016) no.6,  915.

\bibitem{Friedland:2004pp}
  A.~Friedland, C.~Lunardini and C.~Pena-Garay,
  Phys.\ Lett.\ B {\bf 594} (2004) 347
  [hep-ph/0402266].

\bibitem{Miranda:2004nb}
  O.~G.~Miranda, M.~A.~Tortola and J.~W.~F.~Valle,
  JHEP {\bf 0610} (2006) 008
  [hep-ph/0406280].

\bibitem{Ioannisian:2004jk} 
  A.~N.~Ioannisian and A.~Y.~Smirnov,
  Phys.\ Rev.\ Lett.\  {\bf 93}, 241801 (2004)
  [hep-ph/0404060].

\bibitem{Akhmedov:2004rq}
  E.~K.~Akhmedov, M.~A.~Tortola and J.~W.~F.~Valle,
  JHEP {\bf 0405} (2004) 057
  [hep-ph/0404083].

  
\bibitem{Ioannisian:2004vv} 
  A.~N.~Ioannisian, N.~A.~Kazarian, A.~Y.~Smirnov and D.~Wyler,
  Phys.\ Rev.\ D {\bf 71}, 033006 (2005)
  [hep-ph/0407138].

\bibitem{Ioannisian:2015qwa} 
  A.~N.~Ioannisian, A.~Y.~Smirnov and D.~Wyler,
  Phys.\ Rev.\ D {\bf 92}, no. 1, 013014 (2015)
  [arXiv:1503.02183 [hep-ph]].


\bibitem{Akhmedov:2005yt}
  E.~K.~Akhmedov, M.~A.~Tortola and J.~W.~F.~Valle,
  JHEP {\bf 0506} (2005) 053
  [hep-ph/0502154].

\bibitem{Blennow:2003xw} 
  M.~Blennow, T.~Ohlsson and H.~Snellman,
  Phys.\ Rev.\ D {\bf 69}, 073006 (2004)
  [hep-ph/0311098].
  
  

\bibitem{An:2016ses} 
  F.~P.~An {\it et al.} [Daya Bay Collaboration],
  arXiv:1610.04802 [hep-ex].




\bibitem{deHolanda:2004fd}
  P.~C.~de Holanda, W.~Liao and A.~Y.~Smirnov,
  Nucl.\ Phys.\ B {\bf 702} (2004) 307
  [hep-ph/0404042].


\bibitem{Bahcall:1996qv}
  J.~N.~Bahcall, E.~Lisi, D.~E.~Alburger, L.~De Braeckeleer, S.~J.~Freedman and J.~Napolitano,
  Phys.\ Rev.\ C {\bf 54} (1996) 411
  [nucl-th/9601044].


\bibitem{paramet} V. K. Ermilova, V. A. Tsarev and V. A. Chechin, Kr. Soob.
Fiz. [Short Notices of the Lebedev Institute] 5, 26 (1986).
  E.~K.~Akhmedov,
  Sov.\ J.\ Nucl.\ Phys.\  {\bf 47} (1988) 301
   [Yad.\ Fiz.\  {\bf 47} (1988) 475].

\bibitem{Krastev:1989ix}
  P.~I.~Krastev and A.~Y.~Smirnov,
  Phys.\ Lett.\ B {\bf 226} (1989) 341.
  Q.~Y.~Liu, S.~P.~Mikheyev and A.~Y.~Smirnov,
  Phys.\ Lett.\ B {\bf 440} (1998) 319
  doi:10.1016/S0370-2693(98)01102-2
  [hep-ph/9803415].

\bibitem{Acciarri:2015uup}
R.~Acciarri {\it et al.} [DUNE Collaboration],
arXiv:1512.06148 [physics.ins-det].

\bibitem{Alonso:2014fwf}
  J.~R.~Alonso {\it et al.},
  arXiv:1409.5864 [physics.ins-det].


\end{thebibliography}
\end{document}